# First-order quasi-phase-matched blue light generation in surface-poled Ti:indiffused lithium niobate waveguides

A. C. Busacca,[a] C. L. Sones, R. W. Eason, and S. Mailis
*Optoelectronics Research Centre, University of Southampton, Southampton SO17 1BJ, United Kingdom*

We demonstrate efficient first-order quasi-phase-matched second-harmonic generation in a surface periodically poled Ti:indiffused lithium niobate waveguide; 6 mW of continuous-wave blue radiation ($\lambda$=412.6 nm) was produced showing the potential of surface domain inversion for efficient nonlinear waveguide interactions.

[ ]

Efficient generation of coherent blue light is of immediate importance for the development of numerous applications in areas such as displays, optical data storage and printing, and nonlinear frequency conversion achieved by quasi-phase-matched (QPM) interaction remains an attractive route for realizing such devices. QPM requires precise control of periodic domain inversion, with periods that can be as small as $\sim 2$ $\mu$m for first order conversion via second harmonic generation (SHG) from the near-IR into the blue spectral region. Nonlinear crystals such as $LiTaO_3$, $LiNbO_3$, and KTP have been the most commonly used materials for quasi-phase-matching, the choice between them reflecting their respective availability, cost, ease of domain inversion, range of optical transparency and value of nonlinear optical coefficient. Periodically poled lithium niobate (PPLN) however is perhaps the most researched and understood material, and references to PPLN outstrip those on either PPLT or PPKTP by almost an order of magnitude.

SHG results for infrared to visible/blue light have been reported in all of the above electric field periodically poled nonlinear crystals,[1–3] and high efficiencies have been achieved for PPLN in both bulk[4] and waveguide geometries.[5] For the shortest wavelength SHG achievable in lithium niobate first order quasi-phase-matching requires periods of $\sim 2$ $\mu$m, and it is experimentally very difficult to achieve such high aspect ratios in bulk poled material of typical thicknesses $\sim 500$ $\mu$m. The high coercive field required for domain inversion, together with the inherent non-uniformities and defects present in commercially available materials, restrict the routine applicability of electric field poling to periods of the order of $>4$–$5$ $\mu$m in samples of this thickness. To circumvent this problem, different techniques such as controlled spontaneous backswitching[6] and the use of multiple short current pulses[7] have been successfully used to generate periods of the order of 2.2–3.0 $\mu$m in bulk and waveguide geometries, respectively. The aim of the work reported in this letter is to demonstrate the utility of the recently investigated surface poling technique for waveguide geometries for which domain inversion is only required to depths of waveguide dimensions. We have previously demonstrated this technique[8,9] for the fabrication of short period ($\sim 1$ $\mu$m) periodic domain structures in one and two dimensions, and we now report its first implementation in first-order quasi-phase-matched SHG in waveguide structures.

The method of surface domain inversion is based on conventional poling at room temperature[10] with an intentional overpoling step. The intended inverted domain pattern is defined photolithographically on a photoresist layer which has been previously applied (by spin coating) on the surface of the sample while electrical contact is obtained by conductive gel electrodes. The method relies on the observation that upon overpoling there is always a volume of material under the photoresist covered area which will maintain its original polarization state even if the inverted domains have merged into a single domain everywhere else. Large scale, uniform, short period surface poled domains are obtained in this way and periods as short as 1 $\mu$m have been achieved. The depth of these domains is a function of the period of the structure but it is enough to overlap with optical waveguide structures which typically have a depth of a few $\mu$m. Figure 1 shows optical microscopy pictures of a 2.47 $\mu$m period surface inverted domain structure, superimposed on a Ti:indiffused lithium niobate channel waveguide, revealed after HF etching. In Fig. 1 the differentially etched areas can be still identified within the Ti:indiffused channel area. The depth of these surface domains was investigated in Ref. 8 by side HF etching of the $y$ face of the crystal and, more specifically, for domain periods suitable for first-order QPM interactions the domain depth was found to be sufficient for good overlap with waveguide modes (e.g., for 2.47 $\mu$m the mean domain depth is 6 $\mu$m).

An array of Ti indiffused waveguides was fabricated on the $-z$ face of a 500 μm thick $z$-cut congruent lithium niobate substrate supplied by Crystal Technology Inc. (USA). The waveguide fabrication procedure consists of a deposition 100 nm thick titanium stripes of various widths on the sample surface followed by diffusion at 1050 °C in an oxygen atmosphere for 11 h. Several sets of waveguides were fabricated on the same substrate, each set consisting of 12 waveguides separated by 100 μm. The width of the fabricated waveguides within the same set varied from 1.5 to 8 μm. The $-z$ face of the samples was then photolithographically patterned with a 0.7 mm wide and 20 mm long grating structure having a grating period of 2.47 μm and was poled using the method for surface domain inversion. As the pattern was 7 times wider than the separation between consecutive channels several lie in the periodically poled area.

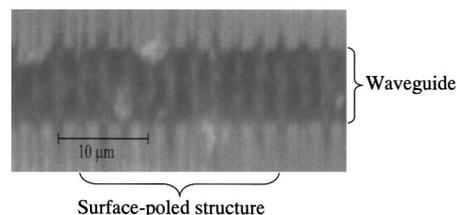

FIG. 1. Optical microscopy pictures of a 2.5 $\mu$m surface periodically poled domain structure superimposed on Ti:indiffused lithium niobate waveguide as revealed after etching in HF acid.

---

[a]Author to whom correspondence should be addressed; electronic mail:

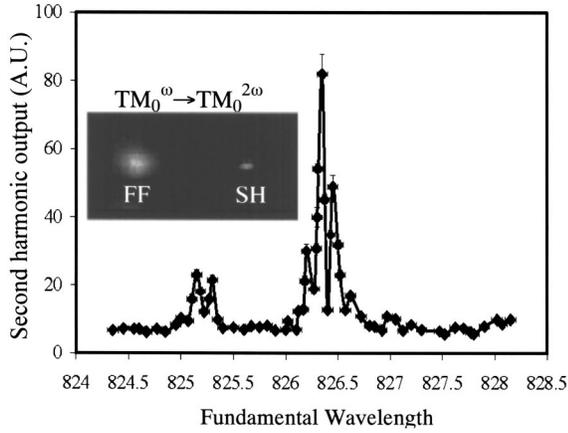

FIG. 2. Wavelength tuning curve for the QPM SHG in a Ti indiffused channel waveguide. The intense peak system corresponds to $TM_0^\omega \rightarrow TM_0^{2\omega}$ coupling scheme while the short wavelength peak system corresponds to higher order waveguide mode coupling. The inset shows the near-field waveguide mode profiles of the fundamental (FF) and second harmonic (SH) waves corresponding to the $TM_0^\omega \rightarrow TM_0^{2\omega}$ interaction.

A titanium sapphire laser was used as a pump source for the QPM SHG experiments. The samples have been end polished so that light could be end coupled using a microscope objective to focus the beam on the polished edge of the waveguide. The output was imaged using a second microscope objective on the surface of the detector and the second harmonic was separated from the fundamental radiation by using a high-pass colored glass filter. During experiments the sample was kept at a temperature of 205 °C in order to avoid possible photorefractive damage generated either by the fundamental or the second harmonic radiation. Efficient SHG was observed on a 2.5 μm wide channel waveguide which was periodically poled with a period of 2.47 μm for TM waveguide mode interaction which utilizes the large $d_{33}$ nonlinear coefficient. Near-field waveguide mode profile monitoring showed that this waveguide could support only the lowest order TM mode in the range of the fundamental wavelengths ($\lambda \sim 825$ nm). A wavelength tuning curve was acquired by measuring the power of the second harmonic as a function of the Ti:sapphire wavelength and is depicted in Fig. 2. In the tuning curve of Fig. 2 two different sets of peaks can be identified, an intense set which is centered at 412.6 nm which corresponds to SHG between the lowest order waveguide mode at the fundamental wavelength ($TM_0^\omega$) to the lowest order mode of the second harmonic ($TM_0^{2\omega}$) and a second set of peaks at shorter wavelengths which corresponds to coupling into higher order waveguide mode at the second harmonic wavelength. The near-field waveguide mode profiles of both the fundamental and second harmonic waves were monitored using a CCD camera. The profiles of the fundamental and second harmonic waves for the $TM_0^\omega \rightarrow TM_0^{2\omega}$ interaction are shown in the inset of Fig. 2. The maximum second harmonic power which was measured, for the $TM_0^\omega \rightarrow TM_0^{2\omega}$ interaction, was 6 mW at 412.6 nm, after correction for Fresnel losses (14%) and filter transmission (70%), for 80 mW of pump power (measured at the exit face of the waveguide and corrected for Fresnel loss), corresponding to an absolute conversion efficiency of 7.4%. As the length of the device is 2 cm the normalized conversion efficiency of the device is 22.8% $W^{-1} cm^{-2}$.

The measured FWHM spectral width of the second harmonic signal is 0.1 nm corresponding to an effective nonlinear grating length of 5 mm, hence only 1/4 of the nonlinear grating length contributes to the frequency conversion process. Although it is difficult to identify the source of the phase mismatch that leads to the reduced effective length this is commonly attributed to irregularities in the waveguide width and/or to temperature inhomogeneities along the sample. Contributing to this assumption is the observed enhanced and asymmetric side lobe structure. All these fabrication imperfections in this nonoptimized device eventually leads to decreased conversion efficiency.[11] From the domain depth profile measurements presented in Ref. 8 it is not expected that the reduced conversion efficiency is due to poor overlap between the waveguide modes and the nonlinear grating but rather due to the nonoptimum mark-to-space ratio of the nonlinear grating. Measurements of the mark to space ratio of surface poled periodic structures performed in side etched (on the $y$ face) samples showed a mark-to-space ratio of ~90% which is far from the ideal 50% value and which would reduce the conversion efficiency by a factor of ~10. If we consider a device length of 5 mm the "effective" conversion efficiency is now 364.8%/W cm². The theoretical value of the conversion efficiency corresponding to a 5 mm grating length with a 90% mark-to-space ratio grating and waveguide propagation loss of 1 dB/cm is 360%/W cm² which is in good agreement with the measured "effective" value. However, all the factors which limit the performance of the device are subject to optimization so that more efficient devices with superior characteristics can be fabricated in the future. This is demonstration showing the potential of such a scheme which utilizes surface inverted domains for efficient first-order QPM waveguide interactions.

We have demonstrated the implementation of the recently developed surface periodic poling method for first-order QPM SHG in Ti:indiffused lithium niobate channel waveguides. The second harmonic power of 6 mW at 412.6 nm was with a normalized conversion efficiency of this nonoptimized device of 22.8%/W cm².

The authors are pleased to acknowledge support from the Engineering and Physical Sciences Research Council (EPSRC) for Research Funding, under Grant No. GR/R47295, and for discussions with Peter G. R. Smith from the ORC, University of Southampton.